\documentclass[%
 aip,
 amsmath,amssymb,
 reprint,%
]{revtex4-1}
\usepackage{xcolor}
\usepackage{hyperref}
\usepackage{graphicx}% Include figure files
\usepackage{dcolumn}% Align table columns on decimal point
\usepackage{bm}% bold math
\usepackage[utf8]{inputenc}
\usepackage[T1]{fontenc}
\usepackage{mathptmx}
\usepackage{etoolbox}
\usepackage{siunitx}
\usepackage{placeins}
\makeatletter
\def\@email#1#2{%
 \endgroup
 \patchcmd{\titleblock@produce}
  {\frontmatter@RRAPformat}
  {\frontmatter@RRAPformat{\produce@RRAP{*#1\href{mailto:#2}{#2}}}\frontmatter@RRAPformat}
  {}{}
}%
\makeatother
\begin{document}

\preprint{AIP/123-QED}

\title[Optical mapping of phases and phase boundaries in nanoconfined fluids]{Optical mapping of phases and phase boundaries in nanoconfined fluids}
% Force line breaks with \\
\author{L. Pierrot Deseilligny}
\email{lauriane.pierrotdeseilligny@trinity.ox.ac.uk}
\author{S. Perkin}%
 \email{susan.perkin@chem.ox.ac.uk}
\affiliation{ 
Department of Chemistry, University of Oxford%\\This line break forced with \textbackslash\textbackslash
}%

\date{\today}% It is always \today, today,

\begin{abstract}

In confined space, deviations from bulk structure and properties are expected due to additional thermodynamic variables. In particular, composition variations arising from surface interactions may lead to additional phases and altered phase transitions. Here, we introduce a non-invasive method for nanoscale composition mapping in confined liquids using the surface force balance (SFB). The method extends conventional SFB analysis from apex measurements to spatially resolved reconstruction of refractive index profiles within confined fluids. When multiple phases are present, the refractive index profiles provide direct access to the position and geometry of the nanoconfined fluid interfaces. We describe the interferogram analysis in detail and establish its range of validity through two model scenarios. First, measurements in air demonstrate the precision of the method and allow detection of a nanometric wetting capillary. Second, we analyse dynamic evaporation of a confined heptane droplet down to \SI{0.1}{pL} volume. The method provides a time-resolved reconstruction of the meniscus geometry throughout the evaporation process, without relying on the Young–Laplace equation. Although evaporation continuously drives the system out of equilibrium, the meniscus remains well described by a catenoidal geometry down to heights of approximately \SI{80}{nm}. At smaller separations, systematic deviations from the catenoidal profile emerge, indicating a crossover from a surface tension-dominated regime to a confinement-dominated regime. Overall, we demonstrate composition profiling as a framework to analyse confinement-induced composition variations and to quantify interfacial thermodynamic effects at the nanoscale. The analysis may be applied to more complex interfacial films and fluids, for example in partially miscible and biomolecular liquid mixtures.\end{abstract}

\maketitle

\section{\label{sec:s1}Introduction}

In natural systems, liquid-liquid phase separations (LLPS) occur in confined geometries, on length scales ranging from the nanometer to the micrometer \cite{Avaro2020,Jin2025,Wang2021}. In confinement, phase stability deviates from bulk predictions, because of the prevalence of two entangled contributions : finite size effects and interfacial effects. Finite size effects act as a constraint on molecular mobility and on the extent of any collective modes or structures. Therefore, while phase growth in bulk is determined by the bulk correlation length $\xi_{c}$ of concentration fluctuations, in confined geometry $\xi_{c}$ is bound by $D$ the smallest dimension of the system. This results in shifting the bulk critical temperature $T_{c}^{\infty}$ to $T_{c}(D)$, with $T_{c}^{\infty}-T_{c}(D)\propto D^{-1/\nu}$ with $\nu$ the critical exponent of $\xi_{c}$ \cite{Binder2010}. Interfacial effects arise from the affinity of the confined species with the surrounding surfaces, and are enhanced in confined geometries because of higher surface-to-volume ratios. If the surfaces are not symmetric with respect to the two components, favouring say a phase $\alpha$ (rich in component $1$) over phase a $\beta$ (rich in component $2$), demixing happens at a potential $\mu_{1}(D)$ lower than at which demixing happens in bulk $\mu_{1}^{bulk}$ \cite{Evans1987}.

Confinement should therefore be treated as a thermodynamic control parameter, not only as a boundary condition. Phase equilibria in confined geometry have been investigated in theoretical and simulation work for simple molecular fluids, however few experimental methods are available to test ideas or to extend towards complex systems. In part this is due to the challenge of directly characterizing spatially-inhomogeneous nanoconfined fluids. For example, nuclear magnetic resonance (NMR) has been applied to distinguish coexisting phases through differences in diffusion and relaxation, but does not yet provide in-situ mapping of the phases \cite{Bramham2022, Murthy2020}. Fluorescence microscopy enables direct spatial mapping of multiphase systems with high resolution, but the addition of fluorescence labels and difficulty of creating controlled confinement limit the ability to resolve quantitative thermodynamic features \cite{Streit2025}.

Here, we present a multiple beam interferometry-based method to characterize phases and phase boundaries of fluids confined down to nanoscale dimensions (sub-pL volumes) between two mica surfaces in crossed-cylinder geometry, and follow their time evolution. To do so, we use the already existing setup of a surface force balance (SFB), but extend the analysis of the interferogram to extract the lateral variations in refractive index and therefore concentration. The theoretical framework and analysis procedure are first described, and then applied to experimental SFB data. First, we determine the range of validity of the method using a calibration in air, with reliable measurements over a confinement range (mica apex separation) of \SI{0}-\SI{120}{nm}. We then study a nanoconfined heptane droplet as it evaporates from \SI{2}{pL} down to \SI{0.1}{pL}. The air–heptane interface is tracked over time, allowing reconstruction of the meniscus geometry throughout the evaporation process, providing direct access to the volume, surface area, and evaporation rate of the fluid. 

\section{\label{sec:s2}Method}
\subsection{\label{sec:ss21}Setup}

\begin{figure}[h!]
    \centering
    \includegraphics[width=\linewidth]{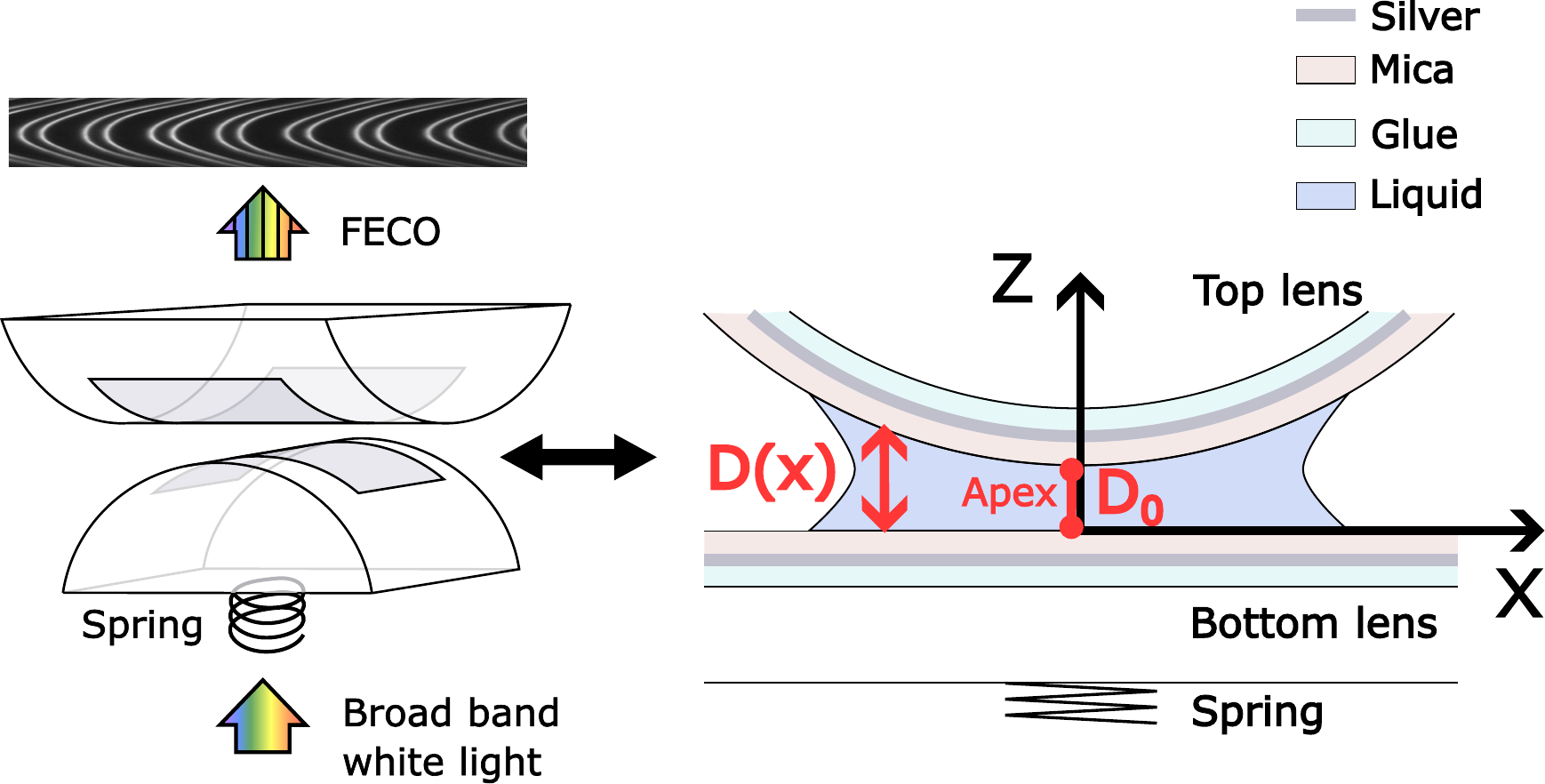}
    \caption{Schematic of the SFB. White light passes through the interferometric cavity built from the crossed-cylinders coated with back-silvered mica. The emerging light is composed of the wavelengths that resonate with the different separations imposed by the cavity. After passing through a spectrometer, the constructive interference emerges as fringes of equal chromatic order (FECO). Under the Derjaguin approximation, the crossed-cylinder geometry is treated as an equivalent sphere-plane geometry. A view from a lateral section is provided on the right, to illustrate the sphere-plane equivalence and to introduce the coordinate system.}
    \label{fig:sfb}
\end{figure}

The Surface Force Balance (SFB) is a well-established technique to measure intermolecular and surface forces in a confined liquid \cite{Hayler2024}. Two hemicylindrical glass lenses are positioned orthogonally and each lens is coated with a back-silvered mica sheet that is a few $\mu$m thick. This setup forms an interferometric cavity, which enables measurement of the separation between the two surfaces. One of the lenses is mounted on a spring of known stiffness which enables measurement of the interaction forces in the fluid \cite{Israelashvili1973}. A schematic of the setup is shown in Figure \ref{fig:sfb}.

Two types of analysis of the interferogram exist: methods based on three-layer interferometry (TLI) \cite{Israelashvili1973,Tadmor2003} and methods based on multi-layer matrix (MLM) \cite{Heuberger2001}. The TLI method operates via an analytical inversion of multiple-beam interferometry. The system is modelled as a symmetric Fabry–Pérot cavity composed of two semi-reflective substrates separated by a homogeneous film of thickness $d$ and refractive index $n$. The measured fringes of equal chromatic order (FECO) are fitted using a phase condition for constructive interference. The separation $d$ and refractive index $n$ are then extracted under assumptions of lateral uniformity, isotropy, and well-defined boundary reflectivities. The TLI method is fast and robust but limited to optically simple systems. The MLM method solves Maxwell’s equations in stratified media. Each layer is described by a propagation matrix, and interfaces are described by Fresnel coefficients. This approach is more general, but involves higher computational complexity and stronger parameter coupling.

While the SFB is widely used to determine the separation and interaction force at the point of closest separation (the apex), it is used less frequently to determine the refractive index of the confined medium. Earlier TLI-based work demonstrated that the refractive index can be extracted as a function of separation by combining fringes of different parity, thereby separating thickness and refractive index contributions \cite{Kekicheff1994}. However, such approaches were generally limited to the apex. Only a very limited number of studies have attempted radial profiling of the refractive index, and these are MLM-based methods \cite{Nowak2008, Kienle}. The work of Góra provides a detailed overview of existing approaches, some of which rely on analysis of the SFB interferogram \cite{Gora12025, Gora22025}. Radial profiling is particularly interesting as it can identify both smooth variations in composition with confinement and sharp changes indicative of phase boundaries. In the following sections we demonstrate how radial profiling of the refractive index can ben achieved using TLI applied to a single interferogram; thus requiring only on a small number of fringes rather than full spectral analysis or multilayer inversion. 

\subsection{\label{sec:ss22}Equations of the three-layer interferometer}

\begin{figure}%[h!]
    \centering
    \includegraphics[width=\linewidth]{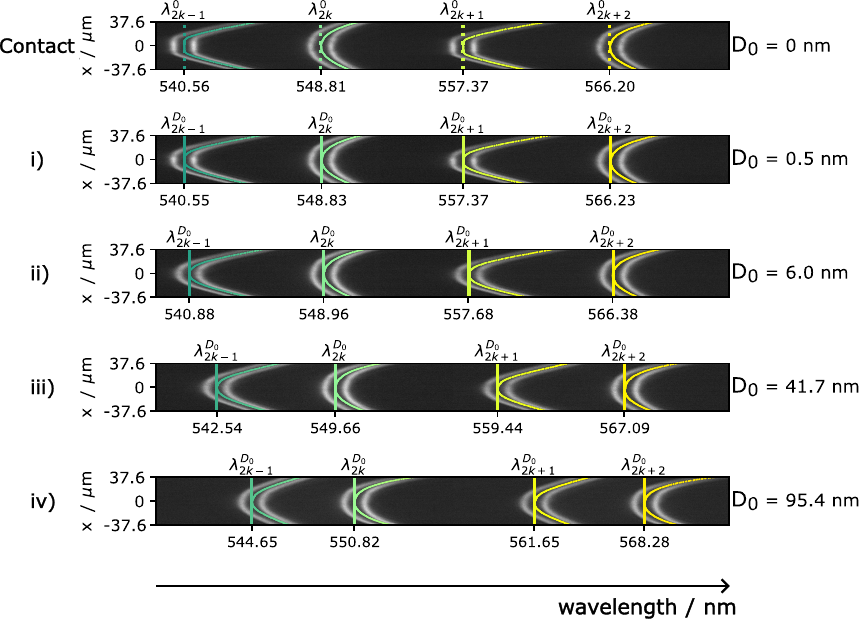}
    \caption{Example FECO interferograms corresponding to different values of the apex separation $D_0$ as indicated on the right. Each interferogram spans a wavelength range encompassing four fringes, each of which is composed of a doublet arising from the mica birefringence. The measured interference appears white, while the fitted centre of mass of each doublet is shown in colour superimposed on the measured data. We denote the wavelength of fringe with order $p$ as function $\lambda_p^{D(x)}$. Vertical lines indicate the apex wavelength of each fringe, i.e. $\lambda_p^{D_0}$. The reference wavelength is $\lambda_p^{D(0)=0}=\lambda_p^{0}$, corresponding to the wavelength of the apex of the fringe where mica sheets are in contact. Note that, for $D_0 = 0$, the angular shape of the left-most fringe indicates the fringe is of odd order, while the subsequent fringe is more rounded, which indicates an even order. The flattening of the fringes at $D_0 = 0$, most apparent on odd order fringes, arises from adhesion between the two mica surfaces which tend to deform the glue above and below.}
    \label{fig:fringes}
\end{figure}

Before any distance measurement, the two mica surfaces are brought into contact. 
The adhesive contact leads to a flattening of the surfaces near the apex, which defines the reference position and fringe order; an example is shown in the top interferogram of Figure \ref{fig:fringes}. Referring to the coordinate system defined in Figure \ref{fig:sfb}, this reference wavelength is $\lambda_p^{D_{0}=0}$ with $D_{0}$ implying the mica separation at the apex, \textit{i.e.} $D_{0}=D(x=0)$. To simplify notation, we identify this reference wavelength as $\lambda_p^{D_{0}=0}=\lambda_p^0$.

The mica surfaces can then be pulled apart, to an apex separation $D_{0}>0$. In Figure \ref{fig:fringes}, example interferograms corresponding to different positions are shown, from $D_{0}=0$ to $D_{0}=$\SI{95.4}{nm}. For any apex position $D_{0}$, the continuous range of separations between the sphere and the plane (obtained for different $x$, as shown in Figures \ref{fig:sfb} and \ref{fig:fringes}) gives rise to a separation profile $D(x,D_{0})$. The constructive wavelength position is $\lambda_p^{D(x,D_{0})}$, with $p$ the order of the fringe, which is an integer. For the sake of brevity, we will write $\lambda_p^{D}=\lambda_p^{D(x,D_{0})}$.

 The birefringent nature of mica can lead to doublets (as observed in Figure \ref{fig:fringes}), or singlets, depending on the relative orientation of the two mica surfaces. In either case, the wavelength of the fringe, $\lambda_p^{D}$, is taken as the center of mass of the singlet or doublet. The open-source \texttt{C++} library \href{https://github.com/micmacIGN/micmac}{\texttt{MMVII}} was used to detect the fringes.

Let $\mu_l^{D}(z)$ be the refractive index field of the confined fluid $(l)$ at separation $D$, as shown in figure \ref{fig:sfb}. 
We define the optical path length:
\begin{equation}
\delta(D) = \int_0^D \mu_l^{D}(z)\,dz
\label{eq:eq1}
\end{equation}

and the mean refractive index:
\begin{equation}
\mu_l^D = \frac{\delta(D)}{D}
\label{eq:eq2}
\end{equation}
Let $\mu_M(\lambda_p^D)$ be the refractive index of mica at wavelength $\lambda_p^D$.

If $p$ is odd, the condition for constructive interference at separation $D$ is \cite{Israelashvili1973, Tadmor2003}: 

\begin{equation}
\begin{array}{rl}
     \tan(K_p \delta(D)) &= \dfrac{2\bar{\mu}_p^D S_p}{(1+(\bar{\mu}_p^D)^2)C_p + ((\bar{\mu}_p^D)^2 - 1)} \\ 
     \tan(K_{p-1} \delta(D)) &= \dfrac{2\bar{\mu}_{p-1}^D S_{p-1}}{(1+(\bar{\mu}_{p-1}^D)^2)C_{p-1} - ((\bar{\mu}_{p-1}^D)^2 - 1)}\label{eq:eq3}
\end{array}
\end{equation}

with

\begin{equation}
\begin{array}{rl}
S_p = \sin\left(\pi \frac{1-\lambda_p^0/\lambda_p^D}{1-\lambda_p^0/\lambda_{p-1}^0}\right),&

C_p = \cos\left(\pi \frac{1-\lambda_p^0/\lambda_p^D}{1-\lambda_p^0/\lambda_{p-1}^0}\right)\\

K_p = \frac{2\pi}{\lambda_p^D}, &
\bar{\mu}_p^D = \frac{\mu_M(\lambda_p^D)}{\mu_l^D}    \label{eq:eq4}
\end{array}
\end{equation}

\subsection{\label{sec:ss23}Refractive index profiling}

The optical path length $\delta(D)$ and the mean refractive index $\mu_l^D$ are treated as unknowns.

Following the standard analysis of multiple-beam interferometry, a single fringe contains only $\delta(D)$. At least two fringes of different parity are therefore required to determine $\delta(D)$ and $\mu_l^D$ independently \cite{Kekicheff1994}.

We rewrite equations \ref{eq:eq3} and \ref{eq:eq4} by introducing $n$ \textit{any} refractive index and defining
$r_p(n) = \frac{\mu_M(\lambda_p^D)}{n}$.

\begin{figure}[h!]
    \centering
    \includegraphics[width=\linewidth]{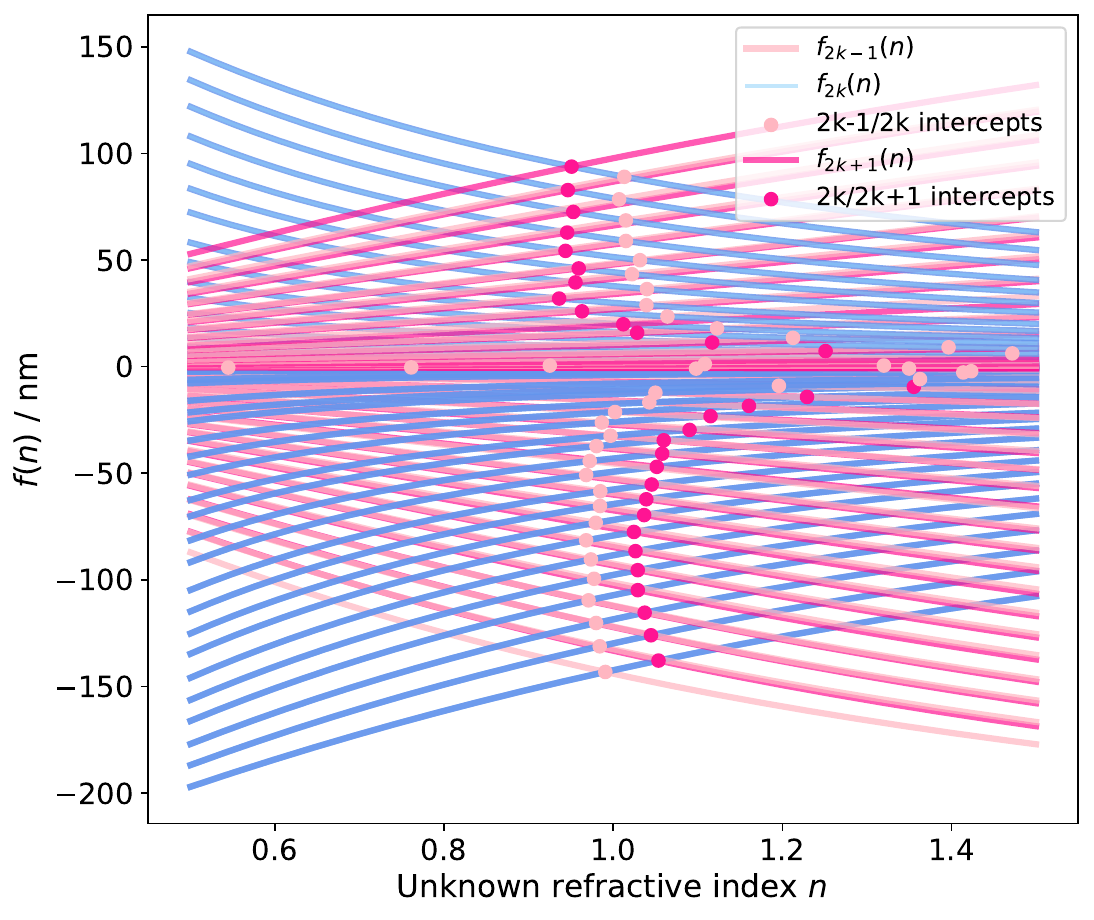}
    \caption{Graphical solution to equations \ref{eq:eq5} and \ref{eq:eq6} applied to the contact interferogram ($D_{0}=0$) from figure \ref{fig:fringes} in order to estimate the optical path length $\delta(D)$ and the mean refractive index $\mu^{D}_{l}$ for all lateral positions $x$. Since the interferogram contains four fringes, for each lateral position $x$, three curves can be calculated: $f^{D}_{2k-1}(n)$, $f^{D}_{2k}(n)$,$f^{D}_{2k+1}(n)$, shown in pink (odd index) and blue (even index). Thus, for this interferogram, there are two intersections defined for each $x$ (shown as dark and light pink points). From the intersections we can compute  estimations of the optical path $\delta(D(x))$ and the refractive index $n(x)$ at each given lateral position $x$.}
    \label{fig:intercepts}
\end{figure}

\begin{figure*}[t]
    \centering
    \includegraphics[width=0.9\textwidth]{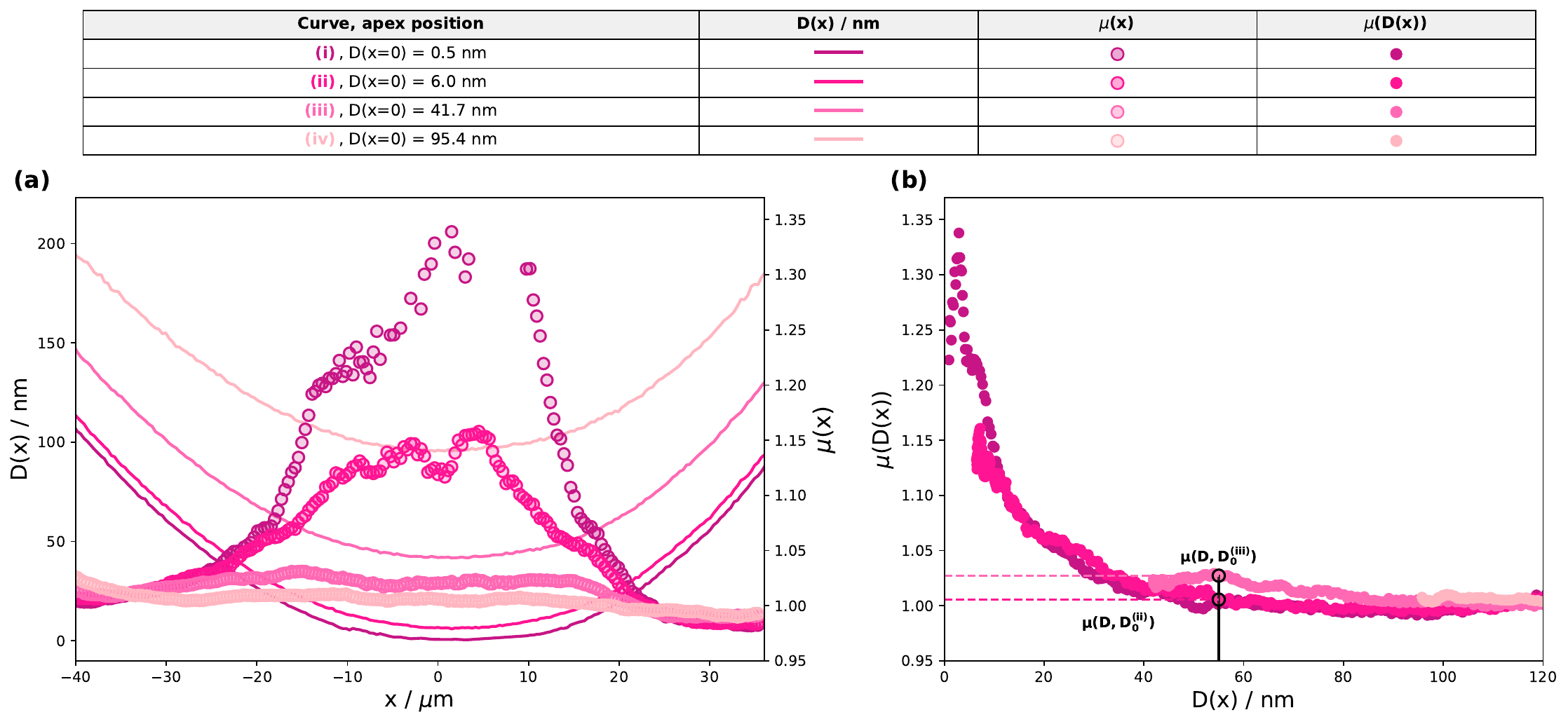}
    \caption{(a) Refractive index and separation profiles in air extracted from the intercepts as in figure \ref{fig:intercepts}, applied to different apex separations which are indicated by the minima of the $D(x,D_{0})$ curves ($D_{0}^{(i)}=$ \SI{0.5}{nm}, $D_{0}^{(ii)}=$ \SI{6.0}{nm}, $D_{0}^{(iii)}=$ \SI{41.7}{nm}, $D_{0}^{(iv)}=$ \SI{95.4}{nm}). In (b) the same data is plotted as refractive index \textit{vs.} separation for each of the four $D_0$ values (four interferograms). }%According to equations \ref{eq:eq5}, if the surrounding medium were at equilibrium in confinement, the surrounding composition would not change, and the local refractive index $\mu_{l}(z)$ would just be a function of the distance to the mica surfaces, not of the apex position $D_{0}$. Here, for a given separation, say $D(x)=$\SI{52}{nm}, the mean refractive index profile for $D_{0}^{(iii)}$=\SI{41.7}{nm} is not the same as for $D_{0}^{(ii)}$=\SI{6.0}{nm}. We explain this increase by the formation of a thin film of water on the hydrophilic mica surfaces because of confinement-induced condensation from the surrounding humidity.}
    \label{fig:airprofile}
\end{figure*}
The optical path can then be expressed as a function of $n$:
\begin{equation}
\begin{array}{rl}
f^{D}_{2k-1}(n) &= \dfrac{1}{K_{2k-1}} \arctan\left( \dfrac{2r_{2k-1}(n)S_{2k-1}}{(1+r_{2k-1}^2(n))C_{2k-1} + (r_{2k-1}^2(n)-1)} \right) \\
\\
f^{D}_{2k}(n) &= \dfrac{1}{K_{2k}} \arctan\left( \dfrac{2r_{2k}(n)S_{2k}}{(1+r_{2k}^2(n))C_{2k} - (r_{2k}^2(n)-1)} \right)
\label{eq:eq5}
\end{array}
\end{equation}

for any integer $k$. For a given separation $D$, both $\delta(D)$ and $\mu_l^D$ are independent of the fringe order. The solution is therefore given by the value of $n=\mu_l^D$ such that:

\begin{equation}
    f^{D}_{2k-1}(\mu_l^D) = f^{D}_{2k}(\mu_l^D) = \delta(D) \label{eq:eq6}
\end{equation}

Graphically, this corresponds to the intersection of the two functions $f^{D}_{2k-1}(n)$ and $f^{D}_{2k}(n)$: the coordinates of this intersection yield $(\mu_l^D, \delta(D))$. This construction requires, for each separation $D$, at least two consecutive fringes of opposite parity, \textit{i.e.} the triplets $(\lambda_{2k-1}^D,\lambda_{2k-1}^0,\lambda_{2k}^0)$ and $(\lambda_{2k}^D,\lambda_{2k}^0,\lambda_{2k+1}^0)$. 

\section{Results and discussion \label{sec:s3}}

 \subsection{Air calibration\label{sec:ss31}}
We begin by applying the above analysis to the interferograms obtained in air (without any added fluid), which provides a foundation for estimating the precision and the range of validity of the method. 

Starting with the contact ($D_0=0$) interferogram of Figure~\ref{fig:fringes}, we first extract the reference wavelengths ($\lambda_{2k-1}^{0},\lambda_{2k}^{0},\lambda_{2k+1}^{0},\lambda_{2k+2}^{0}$) of the four fringes in view. Then, for each $D_0 > 0$ (we take the four examples in Figure \ref{fig:fringes}), we extract for all $x$ the full fringe functions giving $\lambda_{2k-1}^{D(x,D_{0})},\lambda_{2k}^{D(x,D_{0})},\lambda_{2k+1}^{D(x,D_{0})},\lambda_{2k+2}^{D(x,D_{0})}$; we write these as $\lambda_{2k-1}^{D},\lambda_{2k}^{D},\lambda_{2k+1}^{D},\lambda_{2k+2}^{D}$.

 Next, taking the mica dispersive refractive index $\mu_M(\lambda_{p}^D)$ from literature\cite{Bailey_1965}, we compute the functions $f^{D}_{p}(n)$ as in Equation~\ref{eq:eq5} and their intercepts (Equation~\ref{eq:eq6}). An example of the graphical solution is shown in Figure \ref{fig:intercepts}.  In our particular example interferograms (Figure \ref{fig:fringes}), we have four fringes in view which allows us to define three functions: two odd functions $f^{D}_{2k-1}(n)$ and $f^{D}_{2k+1}(n)$, and an even one $f^{D}_{2k}(n)$. We therefore obtain two odd/even intercepts and thus two estimates of the optical path and refractive index at each $x$. In Figure~\ref{fig:intercepts} $\delta(x)$ and $\mu^{D(x)}$ are the ordinate and the abscissa of the intercept points respectively.
 
 By calculating the intercepts for all $x$, one obtains the refractive index profile $\mu^{D(x)}$ as well as the separation profile $D(x)$. We present in Figure~\ref{fig:airprofile} the analysis for each of the four $D_0 >0$ interferograms, i.e. apex separations $D_{0}^{(i)}=$ \SI{0.5}{nm}, $D_{0}^{(ii)}=$ \SI{6.0}{nm}, $D_{0}^{(iii)}=$ \SI{41.7}{nm}, $D_{0}^{(iv)}=$ \SI{95.4}{nm}. In panel (a) we show the refractive index \textit{vs.} lateral position, $\mu(x)$, for each apex distances (dotted profiles). The contact geometry is also extracted, $D(x)$ (solid lines). In panel (b) the same data is represented as refractive index variation with separation, i.e. $\mu(D)$.
 The results return the expected value for the refractive index of ambient air ($T =$\SI{20}{\celsius}) within about 0.03 in the region between \SI{20}{nm} and \SI{120}{nm}. At larger separations ($>$ \SI{120}{nm}), the analysis is limited by phase ambiguities, consistent with the behaviour described by Kienle et al \cite{Kienle}. In addition to verifying general consistency of the method, two features are of interest. First, at small separations, $D<$\SI{20}{nm}, a strong positive deviation in refractive index is detected. Second, it appears that the refractive index is a function of apex separation $D_0$, not only of $D(x)$. We discuss and interpret these observations next. 

 \subsection{Origin of the increase in refractive index near contact: condensation from surrounding humidity \label{sec:ss32}}

\begin{figure}[h!]
    \centering
    \includegraphics[width=\linewidth]{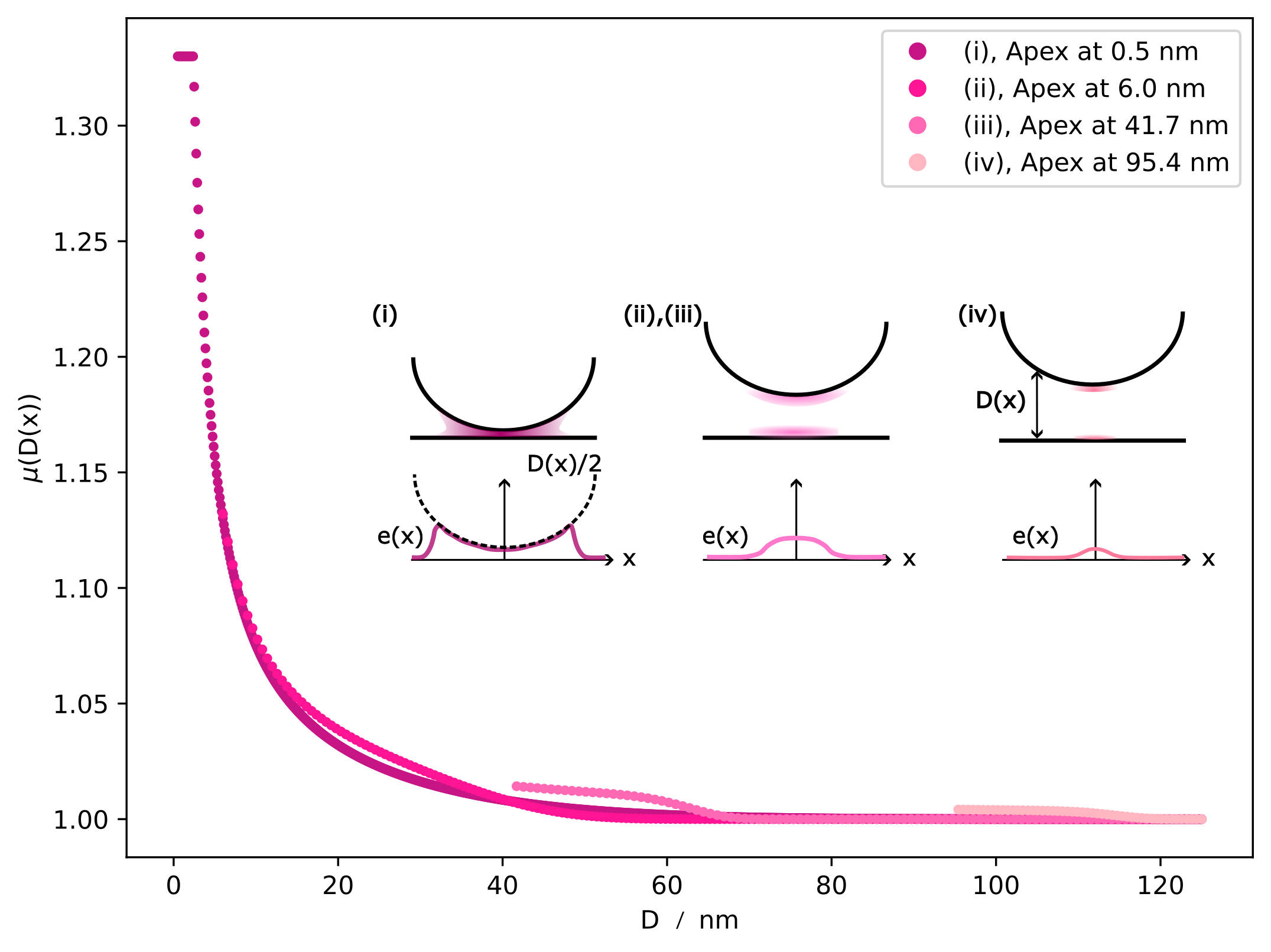}
    \caption{Schematic of the model evolution of $e(x)$ and the associated refractive index profile as a function of separation $D(x)$. The curves and the cartoons (i), (ii), (iii), and (iv) are in direct correspondence with the four curves from figure \ref{fig:airprofile}, corresponding to the apex $D_{0}^{(i)},D_{0}^{(ii)},D_{0}^{(iii)},D_{0}^{(iv)}$. From cartoon (iv) to (iii) and (ii), the water film thickens and spreads, reflected by an increase in parameters $E_{0}$ and $n_{0}$ of equation \ref{eq:eq9}. Cartoon (i) illustrates the situation where the films of each surface have bridged. The overall thickness $2e(x)$ is therefore bound by $D(x)$.}
    \label{fig:cartoonsimu}
\end{figure}  

For the smallest apex separations, $D_0 =$\SI{6.0}{nm} and $D_0 =$\SI{0.5}{nm}, we observe an increase in refractive index of the medium between the mica surfaces detectable out to lateral distances of \textit{ca.}  \SI{25}{\um} and mica separations up to \SI{25}{nm}. 

\begin{figure*}[t]
\centering
\includegraphics[width=0.9\textwidth]{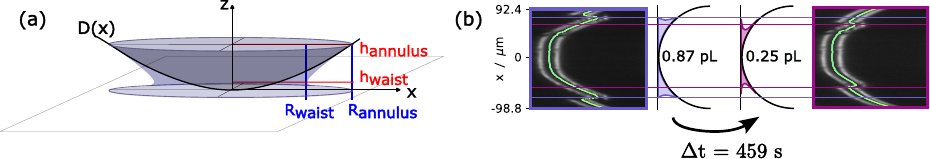}
\caption{Three-dimensional schematic of the heptane annulus wetting on a sphere and a plane. The sphere--plane confinement constrains the annulus to a continuum of separations \(D(x)\). The radius of the annulus is \(R_{annulus}\) and its height is \(h_{annulus}\). The radius of the waist is \(R_{waist}\) and its height is \(h_{waist}\). The air--heptane interface is responsible for a discontinuity in the refractive index which causes the fringes to break at the locus of the meniscus.}
\label{fig:cartoondrop}
\end{figure*}

One possible explanation is numerical instability of the inversion problem arising at small distances, because $f^{D}_{2k-1}(n),f^{D}_{2k}(n) \to 0$ when $D\to 0$. The ordinates of the intercepts that provide the optical thickness also tend to 0 and are rather stable. On the contrary, the abscissas of the intercepts, which result from an inversion, are more and more unstable (see Fig.~\ref{fig:intercepts}). However, it is a clear increase that is observed, rather than increased random fluctuations around a mean value. 

We thus explore the hypothesis that the increase is due to confinement-induced condensation of water on the hydrophilic mica surfaces. On the right-hand side of figure \ref{fig:airprofile}, the profiles (ii) and (iii) correspond to the apex positions $D_{0}^{(ii)}=$\SI{6.0}{nm} and $D_{0}^{(iii)}=$\SI{41.7}{nm}. We observe that for a given distance $D$ the refractive index is not the same for profiles (ii) and (iii):
%1.005 vs 1.027
\begin{equation}
    \mu(D,D_{0}^{(ii)})\ne \mu(D,D_{0}^{(iii)})
    \label{eq:eq7}
\end{equation}

This observed injective behaviour of $\mu(D,D_{0})$ with respect to $D_{0}$ is not anticipated by Equation~\ref{eq:eq4}, or the graphical solution as in Eq.s~\ref{eq:eq5},\ref{eq:eq6}, where the mean refractive index does not depend on the apex position $D_{0}$. Instead, the observed behaviour of $\mu(D,D_{0})$ implies a physical difference in the medium between the surfaces which varies with both $D$ and $D_{0}$. Given that mica is hydrophilic, a feasible explanation is potential condensation of a thin film of water when the surfaces are close to contact arising from higher humidity close to the apex. Since the measured refractive index is a mean over a given separation $D$ (Eq.s~\ref{eq:eq1},\ref{eq:eq2}), the closer the surfaces the more significant is any contribution from a thin surface film to the average. A schematic is provided in figure \ref{fig:cartoonsimu}. 

 Based on the hypothesis that the reservoir of water vapour from the air leads to condensation on the mica surfaces in confinement, we explore a toy model to illustrate the evolution of $\mu(D(x),D_{0})$ if a thin film of water were to grow on the surfaces when $D_{0}$ decreases. 
 
 If a water film of thickness $e$ coats each surface, at a separation $D$, the mean refractive index can be approximated as:

\begin{equation}
    \mu^{D}=\frac{2e\mu_{water}+(D-2e)\mu_{air}}{D}
    \label{eq:eq8}
\end{equation}

% When the surfaces approach, the layer thickness increases in the region near the apex. 
 To model the hypothesis that water vapour condenses increasingly in confinement, we choose to model $e(x)$ as a Gaussian function of $D(x)$ symmetric about $D_{0}$, the apex position. To take into account the film growth with decreasing $D_{0}$, we take its amplitude $E_{0}$ as an decreasing function of $D_{0}$. To model the lateral spreading, we further take its standard deviation $\sigma_{0}$ and its degree $n_{0}$ to be decreasing when $D_{0}$ increases. When the surfaces are close enough, the total thickness $2e(D(x))$ is then bound by the separation $D(x)$, meaning that the two films are in contact (curve (i) of \ref{fig:cartoonsimu}). Thus:

\begin{equation}
e(D(x))= \begin{cases}  E_{0} \exp{\left(-\frac{1}{2}\left(\frac{D(x)-D_{0}}{\sigma_{0}}\right)^{2n_{0}}\right)} &  , D(x)>2e(D(x)) \\
    \frac{D(x)}{2} &,  D(x)\le2 e(D(x))
\end{cases}
     \label{eq:eq9}
\end{equation}

Taking reasonable parameters as in table \ref{tab1}, and fixing $\mu_{water}=1.33$, and $\mu_{air}=1$, the resulting model refractive index profiles under these assumptions are shown in figure \ref{fig:cartoonsimu}.

\begin{table}[h!]
    \centering
    \begin{tabular}{c|c c c c}
          & Curve (i) & Curve (ii) & Curve (iii) & Curve (iv) \\
         \hline
         $D_{0}$ in \unit{nm} & 0.5 & 6.0 & 41.7 & 95.4 \\
         $2E_{0}$ \unit{\angstrom} & 24 & 24 & 18 & 12 \\
         $\sigma_{0}$ in \unit{nm}& 30 & 30 & 20 & 10 \\
         $n_{0}$& 3 & 3 & 2 & 1 \\
    \end{tabular}
    \caption{Chosen parameters  for equation \ref{eq:eq9} in order to evaluate qualitatively the impact of a nanometric film of water on the refractive index profile. The corresponding plots are provided in figure \ref{fig:cartoonsimu}.}
    \label{tab1}
\end{table}

 The simulated curves (Fig.~\ref{fig:cartoonsimu}) are provided for comparison to the measurements (Fig.~\ref{fig:airprofile}(b)), and assist in illustrating the possible qualitative impact of a confinement-dependent water capillary. Close similarity between the model and measurements supports a general interpretation of nanoscale films of water on the mica sheets causing the observed to dependence of $\mu(D(x))$ on $D_{0}$ when $D(x)\to D_{0}$.

\subsection{Heptane evaporation\label{sec:ss33}}

We now demonstrate how refractive index profiling can be used to study a liquid in confined geometry: as a proof-of-concept, we study the evaporation of a heptane droplet. This allows us to track the air--heptane interface during evaporation, revealing the evaporation rate and meniscus radius in droplets down to $\simeq$~10 nm height.

\subsubsection{Annulus and surface geometry\label{sec:sss331}}

A droplet of heptane was injected between the two mica surfaces, which were then brought to molecular contact. The system is opened and heptane evaporates. When the heptane drop has shrunk sufficiently to fit within the field of view of the camera, it has reached a volume of a few pL and can be approximated as an annulus wetting between a sphere and a plane, as shown in figure \ref{fig:cartoondrop}(a). We parametrise the annulus as a surface of revolution $\left(x_{men},y_{men},R_{men}(z)\right)$ about the \(z\)-axis with the origin at the sphere-plane apex and without any particular planar symmetry:

\begin{equation}
    x_{men}^{2}+y_{men}^{2}=R_{men}^{2}(z)
    \label{eq:eq10}
\end{equation}

The FECO spectra were recorded during the evaporation and were analysed as in section \ref{sec:s2}, this time providing the time evolution of the refractive index profile. The presence of the heptane annulus causes a refractive index discontinuity at the air--heptane interface, resulting in a distortion of the fringes as shown in Figure~\ref{fig:cartoondrop}(b). The resulting refractive index profile, $\mu(D)$ as shown in Figure~\ref{fig:freeevap}(a), is sigmoid-like with values close to the refractive index of air ($\mu_{air}=1$) away from the apex and close to the refractive index of heptane ($\mu_{hept}=1.39$) near the apex. The arrow in Figure~\ref{fig:freeevap}(a) shows the front of evaporation receding to the apex over time. 

It is not possible to provide a sharp, non-arbitrary definition of an annulus circumference due to the continuity between the meniscus and any prewetting film. Qualitatively, the edge of the annulus is located around the region where the refractive index plateaus at $\mu_{air}$, while the waist is located where the refractive index profile plateaus around $\mu_{hept}$. Assuming a simplistic model of refractive index similar to equation \ref{eq:eq8}:

\begin{figure*}[t]
    \centering
    \includegraphics[width=0.9\textwidth]{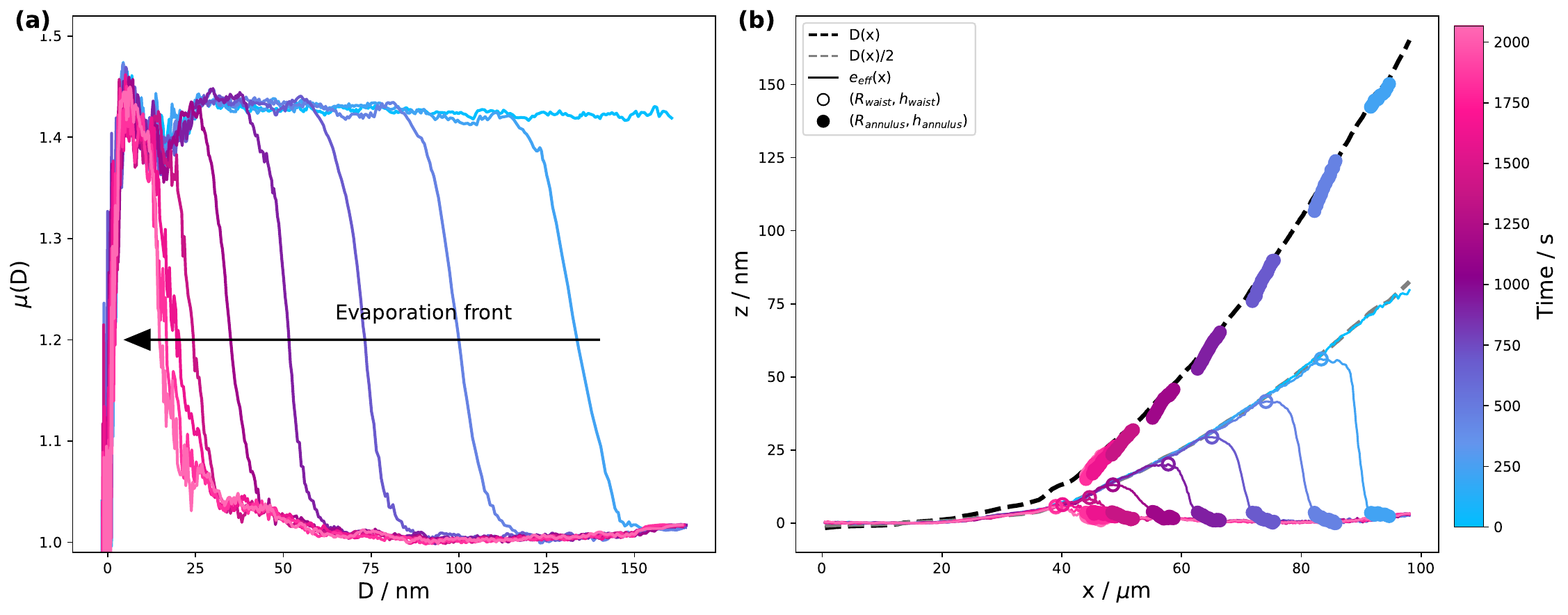}
    \caption{(a): Refractive index as a function of separation for varying time points during the evaporation of the droplet. At large separations the refractive index plateaus to the refractive index of air, while at small separations the refractive index is close to the refractive index of heptane. (b): Time evolution of the effective thickness, \(e_{eff}\), with respect to \(x\), based on the model provided in equation \ref{eq:eq12}. Also shown are the outer edges of the annulus \((R_{annulus},h_{annulus})\) on the sphere and the plane and the locus of the waist \((R_{waist},h_{waist})\).}
    \label{fig:freeevap}
\end{figure*}

\begin{equation}
    \mu^{D}
    =
    \frac{
    2e_{eff}\mu_{hept}
    +
    (D-2e_{eff})\mu_{air}
    }{D}
    \label{eq:eq11}
\end{equation}

where $e_{eff}$ is the effective heptane film thickness, which we can extract from the refractive index profile:

\begin{equation}
    e_{eff}(D)=
    \frac{D}{2}
    \frac{(\mu^{D}-\mu_{air})}
    {(\mu_{hept}-\mu_{air})}
    \label{eq:eq12}
\end{equation}

Two approximations are made: first, the thickness \(e_{eff}\) is an effective thickness because the surrounding heptane vapour contributes to increasing \(\mu^{D}\), so in reality the true liquid thickness is expected to be smaller than the measured value. Second, also arising from the fact that we only access an integrated value of the refractive index (Eqn.~\ref{eq:eq1},\ref{eq:eq2}), is that we cannot rigorously reconstruct any planar asymmetry in the annulus shape, for example arising from gravity. Any asymmetry we observe is therefore likely underestimated compared to reality.

For \(e_{eff}(D)\ll D\), we have \(\left|\frac{de_{eff}}{dD}\right|\ll 1\), and the approximations made above are not prevalent sources of error. Indeed, at such distances it is reasonable to assume that the extent over which the vapour density of heptane accumulates remains small compared to \(D\). Therefore, the regions where \(e_{eff}(D)\ll D\) and \(\left|\frac{de_{eff}}{dD}\right|\ll 1\) can be used to estimate where the meniscus wets the sphere and the plane. In Figure~\ref{fig:freeevap}(b), the filled circles are the regions of \(e_{eff}(D)\) that we attribute to the wetting part of the meniscus.  Then, \(e_{eff}(D)\) is only physically defined when \(e_{eff}(D)\le D/2\). Hence the convergence of \(e_{eff}(x)\) towards \(D(x)/2\) indicates that \(\mu(D)\) no longer varies. We attribute this to the waist of the meniscus. In Figure~\ref{fig:freeevap}(b) the empty circles are the regions of \(e_{eff}(D)\) that we attribute to the waist of the meniscus. Note that \(R_{waist}\) and \(h_{waist}\) denote the radius and height of the waist as defined in Figure~\ref{fig:cartoondrop}.

\begin{figure}[h!]
    \centering
    \includegraphics[width=\linewidth]{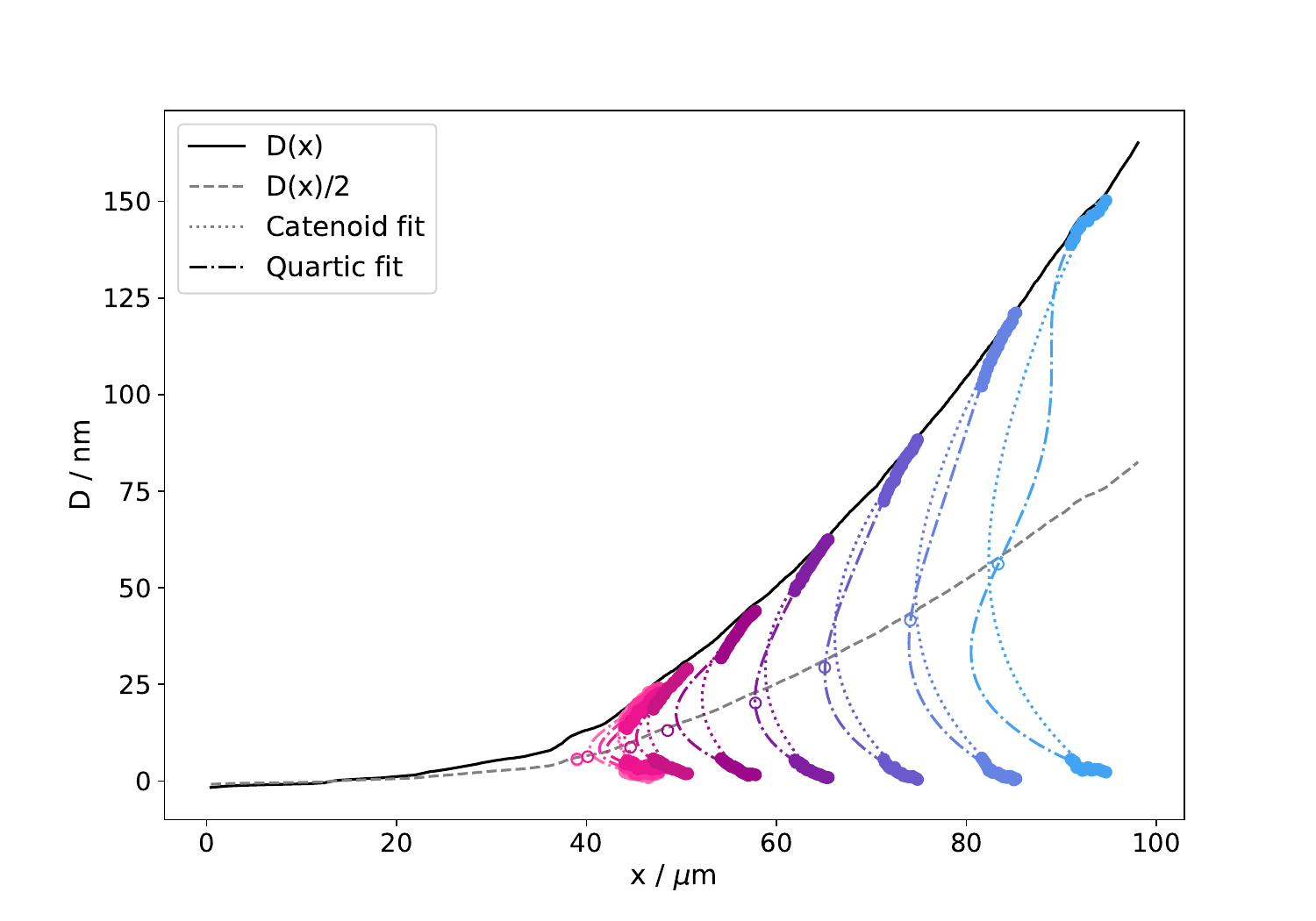}
    \caption{Fitting of the meniscus evaporation surface $S$. The filled circles are points close to the edge of the annulus, while the empty circles are points of the waist of the annulus as depicted in figure \ref{fig:cartoondrop}. We fit those anchor points to reconstruct the meniscus' evaporation surface. Above \SI{80}{nm}, the geometry is well described by a catenoidal curve, but this model collapses below. This deviation from the minimal-surface geometries arises from a strong contraction of the meniscus, showing a crossover to a surface tension dominated regime below \SI{80}{nm}.}
    \label{fig:thicknessfits}
\end{figure}

From the waist of the meniscus and its wetting region extracted from Figure~\ref{fig:freeevap}, we then fit the meniscus surface. Two types of fit were considered: a catenoid surface and a quartic surface, as shown in Figure~\ref{fig:thicknessfits}. We first considered a catenoidal fit since capillary bridges at large separations are expected to minimise surface area due to dominant surface tension\cite{deGennes2004}. The catenoidal fit provides a good representation of the data at distances greater than \SI{80}{nm}. However, at small separations the fit progressively fails at smaller separations. Therefore we introducing a quartic fit to capture the behavior where a stronger contraction of the meniscus waist is observed; the quartic fit performs well in the thinnest films. 

\subsubsection{Evaporation rate\label{sec:sss331}}

\begin{figure*}[t]
    \centering
    \includegraphics[width=0.9\textwidth]{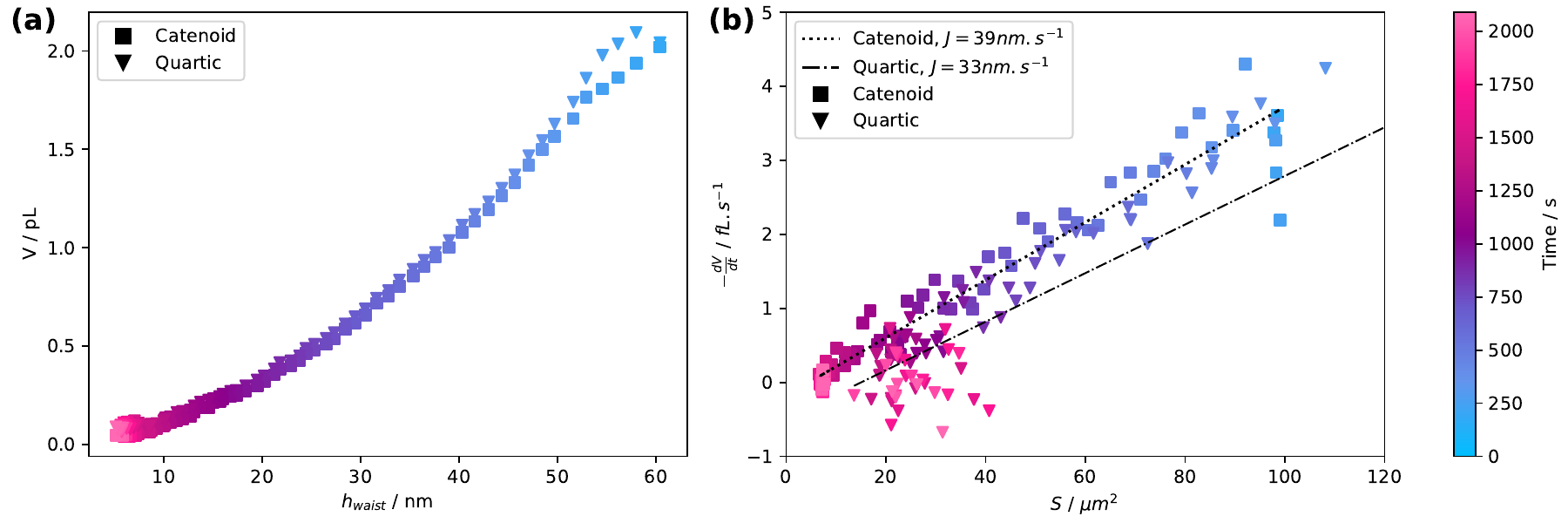}
    \caption{Evolution of the annulus volume and evaporation rate during confined heptane evaporation. The volume is calculated from the reconstructed meniscus geometries. The evaporation rate scales linearly with the annulus surface area, indicating an approximately uniform local evaporation flux over the explored confinement range. With the catenoid fit, we find a flux of \SI{39}{nm.s^{-1}}, while for the quartic fit we find a flux of \SI{33}{nm.s^{-1}}.}
    \label{fig:flow}
\end{figure*}

The reconstruction of \(R(z)\) from equation \ref{eq:eq10} allows us to calculate the volume $V$ of the confined liquid, as shown in figure \ref{fig:flow}, as well as the evaporation surface $S$ of the annulus.

The evaporation rate $-\frac{dV}{dt}$ scales linearly with the annulus surface $S$ over the explored range:

\begin{equation}
    -\frac{dV}{dt}=JS
\end{equation}

This scaling indicates that the local evaporation flux \(J\) remains approximately uniform along the interface, corresponding to a recession velocity of \SI{39}{nm.s^{-1}} with the catenoid fit, and of \SI{33}{nm.s^{-1}} with the quartic fit. The evaporation rate is therefore primarily controlled by the available liquid--air surface area. No significant departure from area-controlled evaporation is observed despite the confinement.

\section{Conclusion \label{sec:s4}}
We introduce a non-invasive interferometric method to reconstruct refractive index profiles in nanoconfined fluids using the SFB geometry. The method combines radial three-layer interferometry with direct analysis of the spatial FECO pattern, enabling nanoscale mapping of composition and interfaces without resorting to fluorescent labelling or multilayer inversion. The reconstruction provides simultaneous access to separation, refractive index, effective film thickness, and meniscus geometry for confining surface separations ranging from molecular diameters up to \SI{120}{nm}. 

Applied to confined heptane evaporation, the method provides a time-resolved reconstruction of the geometry, volume and surface area of an evaporating nanometric meniscus with \SI{0.1}{pL} sensitivity. The evaporating heptane bridge remains close to a catenoidal geometry at large separations, while strong deviations emerge as the waist contracts under confinement. The crossover toward quartic-like profiles indicates the breakdown of the catenoidal description under strong confinement. The evaporation rate scales linearly with the annulus surface, indicating that the local evaporative flux remains approximately uniform despite nanometric confinement. %These observations provide direct experimental evidence that nanoscale confinement modifies the geometry and wetting state of evaporating menisci while preserving area-controlled evaporation kinetics.

Overall, the method establishes a route to direct geometric and compositional characterization of confined fluids from a single interferometric measurement.
Composition, interface geometry, volume and surface area in confined fluids can be determined simultaneously. In future, the method could be applied to characterize composition gradients and phase boundaries in nano-confined multicomponent fluids. For example, it may be relevant for liquid liquid phase separation where confinement and interfacial effects are expected to modify phase stability, composition, and nucleation pathways at nanometric scales. The ability to reconstruct interfaces and concentration profiles simultaneously could enable direct measurements of confinement-induced shifts in coexistence conditions and wetting transitions. More generally, the approach provides a route to experimentally characterize confinement-induced composition variations and phase boundaries in multicomponent fluids.

\begin{acknowledgments}
The authors gratefully acknowledge funding from the UKRI within the FLUXIONIC MSCA doctoral network and from the European Research Council under grant 101001346 ELECTROLYTE. We thank the mechanical and electronic workshops of the Physical and Theoretical Chemistry Department for their technical support with the SFB. We wish to thank the developers of \hyperlink{https://github.com/micmac-V2/MMVII}{\texttt{MMVII}} for implementing the \texttt{ExtractFranges} function for the detection of the fringes.

\end{acknowledgments}

\section*{Data Availability Statement}
The data that support the findings of this study are openly available in the Oxford Research Archive at http://doi.org/10.5287/ora-j2zx5v6zk. The code and a test dataset for the analysis are available in the publically available GitHub repository \hyperlink{https://github.com/cyber-Lolo/InterfeRIOT}{\texttt{InterfeRIOT}}.

\appendix
\section{}

Most of the existing literature on refractive index profiling using SFB and TLI approaches used the linearized equations of the TLI.

\subsection{Linearization \label{app:ssa1}}
\subsubsection{Using the fringes order \label{app:sssa11}}

Papers linearizing the TLI equations usually use the order of the fringes, following the steps detailed below \cite{Israelashvili1973, Tadmor2003, Kekicheff1994}.

Writing $\Delta \lambda^{D}_{p}=\lambda^{D}_{p}-\lambda^{0}_{p}$, the direct linearization of equation \ref{eq:eq3} gives :

\begin{equation}
    \begin{array}{rll}
         \delta(D)=& \frac{p\Delta \lambda^{D}_{p}}{2\bar{\mu}^{D}_{p}}& \text{ for $p$ odd} \\
        \delta(D)=& \frac{(p-1)\Delta \lambda^{D}_{p-1} \bar{\mu}^{D}_{p-1}}{2}& \text{ for $p$ even}
        
    \end{array}
    \label{eq:eq14}
\end{equation}

Leading to : 

\begin{equation}
    \mu^{D}_{l}=\left[\mu_{M}(\lambda^{D}_{p})\mu_{M}(\lambda^{D}_{p-1})\frac{p}{p-1}\frac{\Delta \lambda^{D}_{p}}{\Delta \lambda^{D}_{p-1}}\right]^{1/2}
    \label{eq:eq15}
\end{equation}

Although the simplicity of the formula is appealing, any error in the order of the fringes results in a disastrous estimation of $\mu_{l}^{D}$.
This point is addressed in Kekicheff's paper \cite{Kekicheff1994}, where he insists on the importance of the calibration step, because of the necessity to estimate precisely the fringe orders.

\subsubsection{Linearized formula 2.0 \label{app:sssa112}}

However, it is possible to use a linearized version of the TLI, without the fringes order.

Although in this paper we do not use any linearization, we provide below details of a linearization of the TLI equations that does not use the order of the fringes.

To bypass this difficulty, we can use instead the position of 3 fringes instead of two. This additional information therefore allows to get rid of the order of the fringes which was previously an additional unknown. Since the calculation of the refractive index profile requires three fringes in view anyway, this condition for this linearization to work should not be an additional constraint.

Say the orders of the three fringes are $2k-1,2k,2k+1$.

Applying the linearization to the pair $(2k-1,2k)$ and to the pair $(2k,2k+1)$, one gets :

\begin{equation}
\begin{array}{rl}
    \bar{\mu}_{2p-1}\bar{\mu}_{2p}= & (1-\frac{1}{2p})\frac{\Delta \lambda^{D}_{2p-1}}{\Delta \lambda^{D}_{2p}} \\
    \bar{\mu}_{2p+1}\bar{\mu}_{2p}= &(1+\frac{1}{2p})\frac{\Delta \lambda^{D}_{2p+1}}{\Delta \lambda^{D}_{2p}} 
\end{array}
\label{eq:eq16}
\end{equation}

Which allows us to completely get rid of the orders :

\begin{equation}
     \mu^{D}_{l}=\left[\frac{\mu_{M}(\lambda^{D}_{2p})\Delta \lambda^{D}_{2p}}{2}\left(\frac{\mu_{M}(\lambda^{D}_{2p+1})}{\Delta \lambda^{D}_{2p+1}}+\frac{\mu_{M}(\lambda^{D}_{2p-1})}{\Delta \lambda^{D}_{2p-1}} \right)\right]^{1/2}
     \label{eq:eq17}
\end{equation}

The refractive index calculated from equation \ref{eq:eq17} is way less scattered than the one calculated from equation \ref{eq:eq15}. The results from the linearization of equation \ref{eq:eq17} show good agreement with the non-linearized method detailed in section \ref{sec:ss23}, until \SI{70}{nm}. Above this separation, the linearization does not hold and the calculated refractive index diverges with increasing scatter.

\nocite{*}
\bibliography{aipsamp}% Produces the bibliography via BibTeX.

\end{document}